\documentclass[a4paper,11pt]{article}
\usepackage{pos}
\usepackage{amsmath}
\usepackage{slashed}

\title{On analytic continuation from imaginary to real chemical potential in Lattice QCD}
\ShortTitle{Analytic continuation from imaginary chemical potential}

\author*[a]{Francesco Di Renzo}
\author[a]{Marco Aliberti}
\author[a]{Petros Dimopoulos}

\affiliation[a]{University of Parma \& INFN,\\
  Parco Area delle Scieze 7/A, Parma, Italy}

\emailAdd{francesco.direnzo@unipr.it}
\emailAdd{marco.aliberti@unipr.it}
\emailAdd{petros.dimopoulos@unipr.it}

\abstract{Imaginary baryon number chemical potential simulations are a popular workaround for the (in)famous
sign problem plaguing finite density QCD studies on the lattice. One is necessarily left with the problem of analytically continuing results to real values of $\mu_B$. In the framework of the Bielefeld Parma
Collaboration, we have in recent years studied a multi-point Padé description of the net baryon
number density computed as a function of imaginary baryon number chemical potential. While our
main emphasis has till now been on the determination of Lee-Yang singularities, the method is per
se a natural tool to analytically continue results. We report on the status of our projects with this
respect, comparing the Padé approach to analytic continuation to another, new strategy, which is an application 
of the Cauchy integral formula in the sense of an inverse problem.}

\FullConference{The 41st International Symposium on Lattice Field Theory (LATTICE2024)\\
 28 July - 3 August 2024\\
Liverpool, UK\\}

\begin{document}
\maketitle

\section{The lattice QCD sign problem and the complex $\mu_B$ plane}
\begin{figure}[bh] 
  \centering
  {\includegraphics[width=0.5\linewidth]{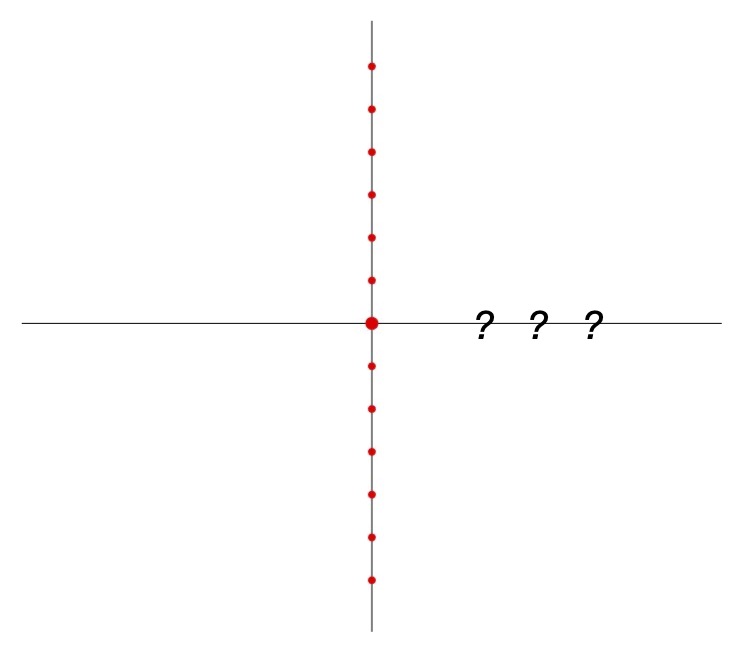}}
  \caption{The complex $\mu_B$ plane. Due to the sign problem, the real ($x$, horizontal) axis is {\em terra incognita}. Instead we can either compute on the imaginary ($y$, vertical) axis or compute Tayor expansions at $\mu_B=0$. In both case, an analytic continuation is due to get physical results.}
  \label{fig:AnaliticLandscape}
\end{figure}

{\em The sign problem is a necessary evil, unavoidable as soon as one integrates out the fermion
fields and expresses the partition function in terms of the gauge fields.} This quotation \cite{deForcrand:2009zkb} is probably one of the most known {\em incipit} of publications on Lattice QCD. Moving to formulas, let's look at what happens when we want to study QCD at finite baryonic density. When a baryonic chemical potential is in place, the Dirac operator satisfies 
\begin{equation*}
    \gamma_5 \,(\slashed{D}+m+\mu_B \,\gamma_0) \,\gamma_5 = (\slashed{D}+m-\mu_B^* \,\gamma_0)^\dagger
\end{equation*}
from which one gets
\begin{equation}
    \det (\slashed{D}+m+\mu_B \,\gamma_0) = {\det}^* (\slashed{D}+m-\mu_B^* \,\gamma_0)
\end{equation}
which in turn implies that there are only two possibilities for the fermionic determinant to be real, {\em i.e.}
\begin{itemize}
    \item $\mu_B=0$
    \item $\mu_B = i \, {\mu_B}_I$
\end{itemize}
For real values of the chemical potential, we end up with a complex weight in the path integral, so that an interpretation in terms of probability fails and the foundation itself of an approach based on Monte Carlo simulations fails as well. All in all, the situation is that depicted in Fig. \ref{fig:AnaliticLandscape}, where we plot a sketch of the complex $\mu_B$ plane: we have access to points on the imaginary axis (including, of course, the origin), but the real axis (where we would like to compute) is {\em terra incognita}. 
The two major solutions to escape the sign problem in lattice QCD are actually based on these two possibilities: one can 
\begin{itemize}
    \item compute Taylor expansions at $\mu_B=0$ \cite{Allton:2002zi,Gavai:2003mf};
    \item compute at imaginary values $\mu_B = i \, {\mu_B}_I$ \cite{deForcrand:2002hgr,DElia:2002tig}.
\end{itemize}
While the first method (the one based on Taylor expansions) is {\em per se} an analytic continutation from zero chemical potential, in the second case one has to explicitly perform an analytic continuation from the imaginary to the real axis in the complex chemical
potential plane. In the following we will be concerned with two methods which somehow aim at
joining the two methods: a certain form of Padè approximants (multi-point Padè) and an application of the Cauchy integral
formula in the form of an inverse problem. \\

This work is in the framework of research performed by the Bielefeld-Parma Collaboration: we computed \cite{Dimopoulos:2021vrk} cumulants of the net baryon number density, given as 
\begin{equation}
    \chi_n^B(T,V,\mu_B)=\left(\frac{\partial}{\partial \hat\mu_B}\right)^n\frac{\ln Z(T,V,\mu_l,\mu_s)}{VT^3}\, ,    
\end{equation}
with $\hat\mu_B=\mu_B/T$, the partition function being that of lattice QCD with $2+1$ flavours in the HISQ regulariztion, with physical value of the pion mass. The cumulants are computed at imaginary values of the baryonic chemical potential and what we have been doing for some time is taking results as inputs for obtaining multi-points Padè - {\em i.e.} rational - approximants \footnote{Till now the main emphasis has been put on obtaining information on the phase diagram by studying the Lee-Yang singularities, which are directly taken from the singularities of the rational function. In particular, the data we will consider in the following are coming from those produced for the study in \cite{Clarke:2024ugt}.}. Having a rational function is {\em per se} a direct way to obtain an analytic continuation, with a very simple recipe: simply take it and compute it for real values of the chemical potential. After briefly discussing this, we will move to yet another method, working again on the same data obtained in the Padè project. The method is a conceptually very simple application of the Cauchy integral formula, which will take us to an inverse problem. Once again, our aim is evaluating an observable (the number density) at real values of the chemical potential taking as inputs computations on the imaginary axis.

\section{Analytic continuation from multi-point Padè}

A Padè approximant is nothing but a rational function $R^{m}_{n}(z)$
\begin{equation}
\label{eq:PadeRatFunct}
R^{m}_{n}(z) = \frac{P_m(z)}{\tilde{Q}_n(z)} = \frac{P_m(z)}{1+Q_n(z)} = \frac{\sum\limits_{i=0}^m \, c_i \, z^i}{1 + \sum\limits_{j=1}^n \, d_j \, z^j}\,.
\end{equation}
The $m$ and $n$ parametrizing the rational function are the degrees of the polynomials at numerator and denominator respectively.
Notice that the rational function depends essentially on $n+m+1$ parameters. \footnote{In principle we should demand that there is no (common) zero of both numerator and denominator. In practice, we cannot exclude the possibility of common zeros, 
and we will instead live with those, which are even a quite common event.}
The main idea is having this function as a {\em smart proxy} for another function $f(z)$ we are really interested in, for which we typically have a limited amount of information. $R^{m}_{n}(z)$ is basically intended for
\begin{itemize}
    \item interpolating $f(z)$;
    \item extrapolating $f(z)$ beyond the region in which we (at least partially) know it;
    \item hunting for the singularities of $f(z)$.
\end{itemize}
While a polynomial approximation of $f(z)$ could be in principle as good as a rational function with respect to the first two points, (hints on) singularities are a piece of information which we would miss with polynomials.\\
The by far more common form of Padè approximants is what is named {\em single point} Padè. We will instead be concerned with 
{\em multi-points} Padè, which is natural to consider when we know a few Taylor expansion coefficients of our
function $f(z)$ at different points \footnote{It is clear that the number of coefficients we know can be different
at different points. For the sake of simplicity we will however assume that $f^{(s-1)}$ is the highest order derivative which is known at each point (together with all derivatives of degree $0 \leq g <s-1$).}, {\em i.e.}
\begin{equation}
\ldots, \, f(z_k), \, f'(z_k), \ldots, f^{(s-1)}(z_k), \; \ldots \; \;
k=1 \ldots N 
\label{eq:OURdata}
\end{equation}
Since we want $R^{m}_{n}(z)$ to be a good interpolation for $f(z)$, it is natural to require that
$$ \left(\frac{d}{dz}\right)^g R^{m}_{n}(z)|_{z=z_k} = f^{(g)}(z_k) \,.$$
The somehow simplest case we can discuss is that of having $n+m+1=Ns$,
in which case we can solve a linear system 
\begin{equation}
\label{eq:LinearProblem3}
\begin{split}
 & \vdots \\
P_m(z_k) - f(z_k)Q_n(z_k) &= f(z_k) \\ 
P_m'(z_k) - f'(z_k)Q_n(z_k) - f(z_k)Q_n'(z_k) &= f'(z_k) \\ 
 & \vdots \\
P_m^{(s-1)}(z_k) - f^{(s-1)}(z_k)Q_n(z_k) - \hdots -f(z_k)Q_n^{(s-1)}(z_k)
&= f^{(s-1)}(z_k) \\
 & \vdots \\
\end{split}
\end{equation}
The solution of this system of linear equations returns the coefficients of the polynomials $P_m$ and
$Q_n$. We could of course rely on different methods to get these coefficients, all somehow
related to the idea of minimizing a generalized $\chi^2$. In practice, 
we could minimize the distance between the input Taylor 
coefficients and the relevant rational function, weighted by the 
errors on the input coefficients (the latter will in our case
come from Monte Carlo measurements). Notice that this is
equivalent to solving an over-constrained system ($n+m+1<Ns$) in a least
squares sense (We compared the latter method to the linear solver method 
in \cite{Dimopoulos:2021vrk}).\\

\begin{figure}[h] 
  \centering
  {\includegraphics[width=1.0\linewidth]{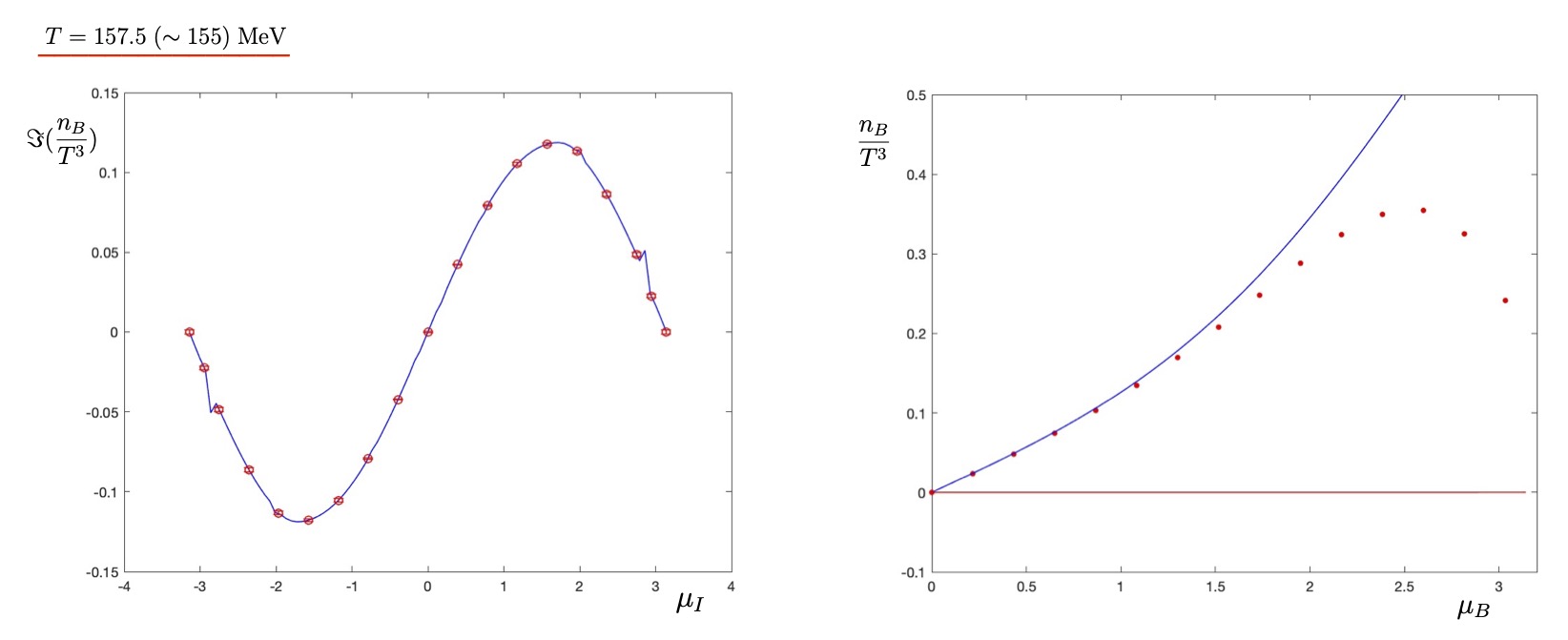}}
  \caption{Multi-point Padè approximants for the number density at imaginary chemical potential (left; this is an interpolation of data from Monte Carlo) and (analytically continued) at real chemical potential (right; this is instead an extrapolation). In the right panel, we plot for comparison the sum of the Taylor expansion up to the eight order (red dots). As explained in the text: the analytically continued results are stable up to $\hat{\mu}_B \sim 1.5-2.0$.}
  \label{fig:AnalitiContPADE}
\end{figure}

Figure \ref{fig:AnalitiContPADE} is an example of how to get an analytic continuation from a multi-point Padè: we present results 
for the number density. In simple words: we will look at the rational function on the left hand side as an interpolation, on the left hand side as an extrapolation. In the left panel, we plot the original results at imaginary chemical potential as red circles, together with the (interpolated) Padè rational function, which is the blue line. Even a casual reader can notice that errors are hardly visible for the original points and (in practice) negligible for the interpolation. A more careful reader will notice a few small {\em bumps} in the rational function. These do not come as a surprise: they are the result of a non-perfect cancellation of some zeros of numerator and denominator. In practice, not all the zeros and poles of our rational function are genuine information accounting for zeros and poles of the number density. In the right panel, we plot again our rational function, this time computed for real values of the chemical potential (blue line). For comparison we also plot (red dots) the sum of the Taylor series up to the eight order, as taken from (\cite{Bollweg:2022rps}). Beware! Errorbars are {\bf not} displayed (we are mainly concerned with {\em trends}). This has been obtained at a given temperature ($T\sim 155 MeV$) at fixed cutoff. Discrepancy beyond $\hat{\mu}_B \sim 1.5-2.0$ are not to be really taken as much significant. The main point is that, while at imaginary chemical potential we have little dependence of the rational function on the interval in which we take inputs for playing the Padè game, on the real axis this dependence can be very much significant. All in all, we have different behaviors beyond the $\hat{\mu_B} \sim 1.5-2.0$ threshold, depending on the input we take for the Padè approximant. We should nevertheless notice that in the same region, also errorbars on the sum of the Taylor series are huge. We can summarize in this way: while {\em (a)} analytic continuation of a (Padè) rational function is in principle trivial business, nevertheless {\em (b)} beyond a given threshold, we have a large dependence of results on the input for the Padè machinery \footnote{{\em e.g.}, number of derivatives we take into account, imaginary chemical potential interval we select to start with, and so on.}; also, {\em (c)} this threshold is the same that separates the region in which the errors of Taylor series sums are small from that in which they are large.

\section{Analytic continuation as an inverse problem (Cauchy integral formula)}

\begin{figure}[t] 
  \centering
  {\includegraphics[width=1.0\linewidth]{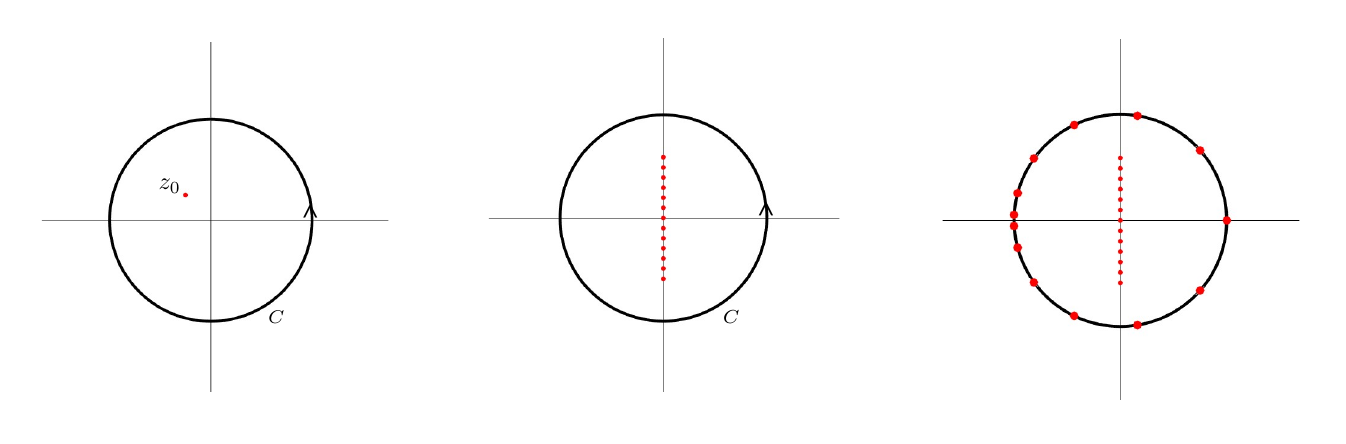}}
  \caption{Pictorial representation of our numerical interpretation of the Cauchy integral formula as an inverse problem (see text).}
  \label{fig:CauchyFormula}
\end{figure}

Figure \ref{fig:CauchyFormula} is a {\em cartoon} of a fundamental result in analytic functions, {\em i.e.} the Cauchy (integral) formula. We consider a function $f(z)$, defined in a domain $D$ of the complex plane, analytic everywhere within and on a simple closed contour $C$ (taken in the conventional positive sense), and a generic point $z_0$ inside $C$. We have that 
\begin{equation}
    f(z_0) = \frac{1}{2\pi i}\, \oint_{C} \dfrac{f(z)}{z-z_0}\, dz
\label{eq:CauchyFormula}
\end{equation}
(look at left panel of Figure \ref{fig:CauchyFormula}). All in all, for a function $f$ which is analytic on a contour $C$ and in its interior, all values of $f$ inside $C$ are entirely determined by its values on the contour.
The Cauchy integral formula can of course be applied to obtain values of $f(z)$ on the imaginary axis (now look at the central panel of Figure \ref{fig:CauchyFormula}). If we consider the baryonic chemical potential complex plane and we take for $f(z)$ the number density, these are just values we can compute by Monte Carlo simulations (there is no sign problem).  \\
A convenient contour $C$ is a circle of radius $R$ centered in $z=0$, for which 
\begin{equation}
    f(z_0) = \frac{1}{2\pi}\, \int_{0}^{2\pi} \frac{f(R\,e^{i\theta})\, R\, e^{i\theta}}{R\, e^{i\theta}-z_0}\, d\theta\, .
\label{eq:CauchyFormula2}
\end{equation}
In this way the Cauchy integral formula is expressed by an integral on the real axis, which can be numerically computed in a convenient quadrature scheme ({\em e.g.} via Gauss-Legendre quadrature) \footnote{We assume the reader is familiar with this result of numerical analysis: the computation of a real integral is traded for the computation of the sum of products of values of the integrand computed at {\em nodes} times corresponding {\em weights} ($w_k$).} 
\begin{equation}
f(z_0) \simeq \frac{1}{2\pi}\, \sum_{k=1}^{n} w_k\, \frac{f(R\,e^{i\theta_k})\, R\, e^{i\theta_k}}{R\, e^{i\theta_k}-z_0}\, .
\label{eq:CauchyFormulaGL}
\end{equation}

\begin{figure}[t] 
  \centering
  {\includegraphics[width=1\linewidth]{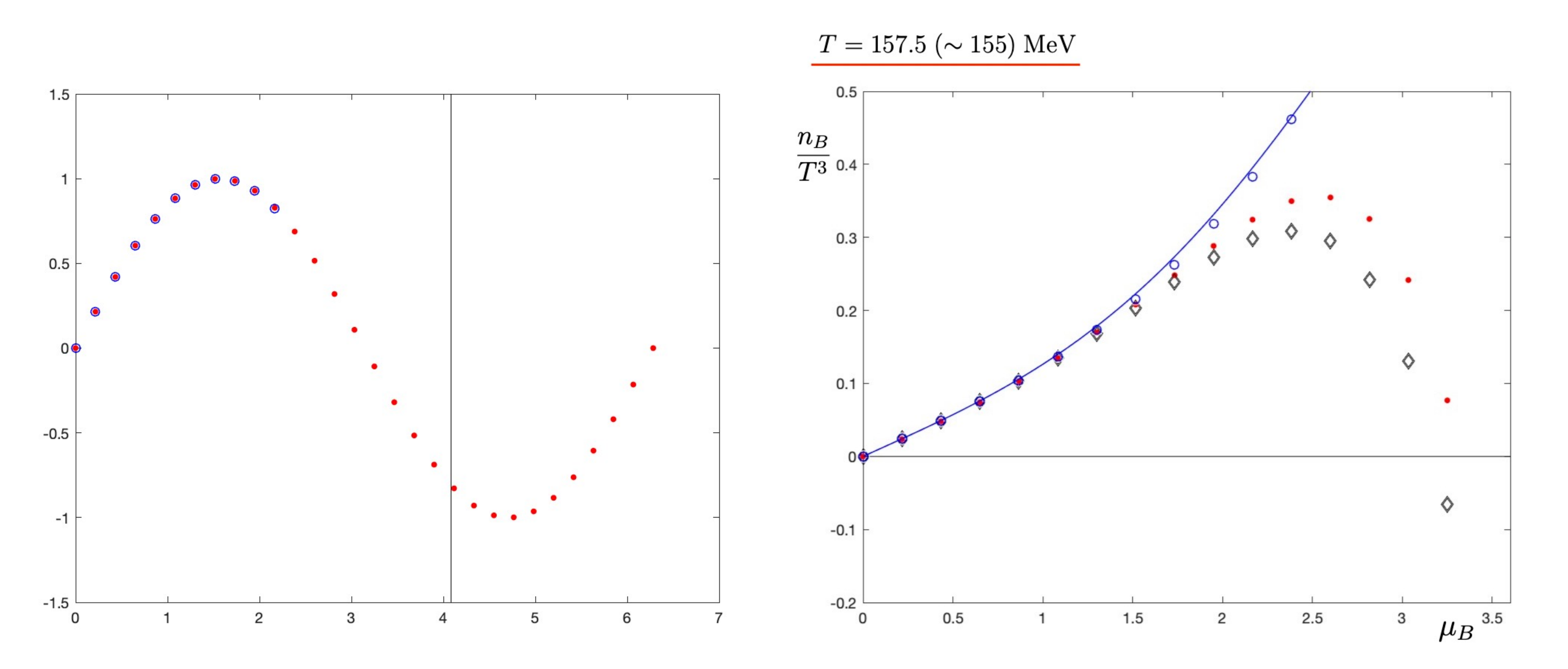}}
  \caption{Our inverse-problem-procedure at work. Left panel: taking inputs on the imaginary axis, we compute values of the $sin$ function on the real axis (the vertical line is the threshold we cannot trespass, {\em i.e. }the radius $R$ in our computation). Right panel: application to lattice QCD (see text).}
  \label{fig:comboFINAL}
\end{figure}

We now proceed to get an {\em inverse problem} out of this. As a result of {\em e.g.} Monte Carlo computations, suppose we know a finite set of values of our function $f(z)$, {\em i.e.} $\{f(z_i) \equiv y_i\, |\, i=1,2, \ldots, n \}$ at a respective set of points. It should be clear what we want to do: the $z$-plane will be the complex chemical potential plance, $f(z)$ will be the number density and the points we want to consider will be right on the imaginary axis (again, where Monte Carlo works). If we can  assume that our function $f(z)$ is analytic on and inside a circle that is centered in the origin and has radius $R$, we can write the Gauss-Legendre quadrature formula of Eq.~(\ref{eq:CauchyFormulaGL}) as (notice that now we trust the formula as exact!)
\begin{equation}
y_i = \frac{1}{2\pi}\, \sum_{k=1}^{n} w_k\, \frac{R\, e^{i\theta_k}}{R\, e^{i\theta_k}-z_i}\, \hat{f}_k\, , \,\, i=1,2, \ldots, n\, ,
\label{eq:CauchyFormulaGLbis}
\end{equation}
with $\hat{f}_k = f(R\, e^{i\theta_k})$ (now look at the right panel of Figure \ref{fig:CauchyFormula}).
And here comes the inverse problem: we consider the previous relation Eq. (\ref{eq:CauchyFormulaGLbis}) as a linear system which we want to solve 
\begin{equation}
A\, \mathbf{x} = \mathbf{b} \, \Leftrightarrow\, \mathbf{x} = A^{-1}\, \mathbf{b} \,.
\label{eq:matrix_eq}
\end{equation}
$A$ is an $n \times n$ matrix with elements 
$$
A_{ik} = \frac{1}{2\pi}\, w_k\, \frac{ R\, e^{i\theta_k}}{R\, e^{i\theta_k}-z_i}\, 
$$
while $b_i = y_i$ and $x_k = \hat{f}_k$. By solving the linear system we get a number of values (at a number of points) of our function on the contour $C$. This has come from our knowledge of values of $f(z)$ (at a number of points) in the interior of $C$ (namely, these are points on the imaginary axis). If we think in terms of our original Eq. (\ref{eq:CauchyFormula}), this is an inverse problem. \\
Knowing the values of our function at the nodes which are relevant for the numerical version of the Cauchy integral formula - Eq. (\ref{eq:CauchyFormulaGLbis}) - the latter can be used in a {\em direct} (as opposed to {\em inverse}) way: we will compute values of $f(z)$ in other points in the interior of $C$, in particular on the real axis. This is the analytic continuation we are interested in: we get unknown values on the real axis from known values on the imaginary axis. 
Notice that there is a version of the integral Cauchy formula for derivatives, namely for the $n$-th one
\begin{equation}
f^{(n)}(z_0) = \frac{n!}{2\pi i}\, \oint_{C} \frac{f(z)}{(z-z_0)^{n+1}}\, dz \, .
\label{eq:CauchyIF_der}
\end{equation}
If we take $z_0=0$,  for a given parity we have pieces of information {\em for free}: indeed we can profit from this. \\
In Figure \ref{fig:comboFINAL} one can see this machinery at work. In the left panel, one can inspect the result of the computation of values of the $sin$ function on the real axis from the knowledge of values on the imaginary axis.
We would say we are doing somehow well, {\em i.e.} the method appears to work: can we trust this? In other words: why could the method fail? A failure is likely because {\em (a)} the numerical version (via Gauss-Legendre quadrature) of the Cauchy formula is {\em not exact} and {\em (b)} the linear system is in general {\em ill-conditioned}. The combination of these two points can result in a failure. Actually, at the time of the conference, if we inspected the obtained values of $\hat{f}_k$, they looked like {\em non-sense}. Still, there was the success depicted in Fig. \ref{fig:comboFINAL} and this could be explained saying that we had an {\em effective quadrature} of our own. Performing other tests, we could indeed provide some pieces of evidence for this interpretation. \\
In Fig. \ref{fig:comboFINAL} (right panel) we also plot the method at work for lattice QCD, {\em i.e.} indeed we went for the analytic continuation of results obtained on the imaginary axis for the number density. As in Fig. \ref{fig:AnalitiContPADE}, the blue solid line is the analytic continuation we got by our (multi-point) Padè approximant and the red dots are the result of summing the Taylor series (up to order eight). Blue circles and black diamonds are both obtained by the inverse-problem Cauchy formula. They differ from each other: actually we took different inputs to solve the linear systems in Eq. (\ref{eq:matrix_eq}). Errorbars are once again not plotted. Notice that, at the time of the conference, we got just the same indetermination we mentioned at the end of sec. 2: results changed if we changed the input for our procedure. Blue circles results are very close to Padè results and indeed the input regions used for the two methods were in these cases close to each other. This dependence on input data was once again showing up for values of $\hat{\mu}_B$ beyond a threshold at $\hat{\mu}_B \sim 2$. \\
Finally, we notice that the inverse-problem machinery we described has an obvious other application for Laplace (anti)transforms. The relevant quadrature formula is in this case Gauss-Laplace. There is a variety of possible applications for this (spectral functions, but not only that). \\ 

We reported the status of our studies at the time of the conference. Since then, we made substantial progress, both for the Cauchy formula which is relevant for the subject of this work and for the Laplace transform case. These will be accounted for in a publication we will release soon.

\acknowledgments
\noindent
This work is supported by INFN under the research project {\em i.s. QCDLAT}. It is our pleasure to thank all our colleagues in the Bielefeld-Parma Collaboration: we plan to apply all this machinery in the context of our common research plans.
\newpage

\bibliographystyle{JHEP}
\bibliography{MYbiblio.bib}

\end{document}